\documentstyle[twoside,fleqn,espcrc2,epsf]{article}

\newcommand{\fbstat}{f_B^{static}}
\newcommand{\fb}{f_B}
\newcommand{\fds}{$f_{D_s}$}
\newcommand{\mystrut}{\vrule height 12pt depth 5pt width 0pt}
\newcommand{\shortstrut}{\vrule height 11pt depth 4pt width 0pt}

\newcommand{\AmS}{{\protect\the\textfont2
  A\kern-.1667em\lower.5ex\hbox{M}\kern-.125emS}}

\title{Heavy-light decay constants---MILC results with the Wilson action\footnotemark}
\author{The MILC Collaboration}
\author{ Claude~Bernard,\hskip-0.03in
\address{Department of Physics, Washington University, St.~Louis, MO 63130, USA} 
Tom~Blum,\hskip-0.03in
\address{Department of Physics, Brookhaven National Lab, Upton, NY 11973, USA} 
Thomas~A.~DeGrand,\hskip-0.03in
\address{Physics Department, University of Colorado, Boulder, CO 80309, USA} 
Carleton~DeTar,\hskip-0.03in
\address{Physics Department, University of Utah, Salt Lake City, UT 84112, USA}
Steven~Gottlieb,\hskip-0.03in
\address{Department of Physics, Indiana University, Bloomington, IN 47405, USA}
Urs~M.~Heller,\hskip-0.03in
\address{SCRI, Florida State University, Tallahassee, FL 32306-4052, USA} 
Jim~Hetrick,$\,\null^{\rm a}$
Craig~McNeile,$\,\null^{\rm d}$
Kari~Rummukainen,\hskip-0.03in
\address{Universit\"at Bielefeld, Fakult\"at f\"ur Physik, Postfach
100131, D-33501 Bielefeld, Germany} 
A.~Soni,\hskip-0.03in$\,\null^{\rm b}$
Bob~Sugar,\hskip-0.03in
\address{Department of Physics, University of California, Santa Barbara, CA 93106, USA} 
Doug~Toussaint,\hskip-0.03in
\address{Department of Physics, University of Arizona, Tucson, AZ 85721, USA} 
and Matthew~Wingate$\,\null^{\rm c}$
} 

\begin{document}

\begin{abstract}
We present the current status of our ongoing calculations of pseudoscalar
meson decay constants for mesons that contain one light and one heavy
quark ($f_B$, $f_{B_s}$, $f_D$, $f_{D_s}$).  
We are currently generating new gauge
configurations that include dynamical quarks and calculating the decay
constants.  In addition, we have several new results for the static
approximation.  Those results, as well as several refinements to the
analysis, are new since Lattice '96.  Our current (still preliminary)
value for $f_B$ is
$156\pm 11 \pm 30 \pm 14$ MeV, where the first error is from statistical and
fitting errors, the second error is an estimate of other systematic errors
within the quenched approximation and the third error is an estimate of the
quenching error. For the ratio
$f_{B_s}/f_B$, we get  $1.11\ \pm 0.02\ \pm 0.03\ \pm 0.07$. 

\end{abstract}

\maketitle

\section{INTRODUCTION}
\renewcommand{\thefootnote}{\fnsymbol{footnote}}
\footnotetext{Invited talk presented by S.~Gottlieb at ``Lattice QCD on 
Parallel Computers,'' University of Tsukuba, March, 1997, to appear in the
proceedings.}

Heavy-light meson decay constants such as $f_B$ and $f_{B_s}$ are of great
interest because they are necessary to interpret
current and future measurements of $B$-$\bar B$ and $B_s$-$\bar B_s$ mixing.  
Only with knowledge of the decay constants and
corresponding $B$-parameters can we 
extract the Cabibbo-Kobayashi-Maskawa matrix elements and 
begin to decipher the fundamental 
information about the Standard Model (or physics beyond it)
that is encoded in the CKM matrix.
Recent reviews of this active field can be found in Ref.~\cite{REVIEWS}.
Here we will concentrate only on the recent work of the MILC Collaboration.

We have been involved in the calculation of pseudoscalar meson decay
constants for some time now \cite{US}.  Our initial goal was to calculate $f_B$
in the quenched approximation.  This calculation was done using gauge
couplings between 5.7 and 6.52.  Our current goal is to estimate the
error that comes from the quenched approximation; however, we have also
been refining our analysis.

To estimate the quenching error, we have calculated decay constants on lattices
generated with two flavors of dynamical Kogut-Susskind quarks.  The gauge
coupling ranges from 5.445 to 5.7 and a variety of quark masses have
been explored.  (Table \ref{tab:lattices} summarizes the lattices that have
been used in our calculations.)
To go from an estimate of the quenching error to an actual
calculation with dynamical quarks will require much work.  It will be
necessary to extrapolate in the dynamical quark mass to the $u$/$d$ mass.
It will also be necessary to extrapolate results to zero lattice
spacing.  The dynamical strange quark, which does not appear in our
calculations, will have to be included.  Also, it would be nice to use
the same lattice formulation of the dynamical and valence quarks.
At the coupling $6/g^2=5.5$, we have results for four quark masses.  We
plan to have the same number at 5.6, so we will be able to attempt
extrapolation in the light quark mass for at least two couplings.
We have no current plans to include the effects of a dynamical strange
quark.

\begin{table}[tb]
\newlength{\digitwidth} \settowidth{\digitwidth}{\rm 0}
\catcode`?=\active \def?{\kern\digitwidth}
\caption{Lattices used in our calculation. (Lattice Q is used only to
study finite volume effects and is not included in the final results.)}
\label{tab:lattices}
\begin{tabular}{lccc}
\hline
\multicolumn{4}{c}{\mystrut Quenched}\\
\hline
\mystrut Name&$6/g^2$&size&\#\\
\hline
A${}^{\dagger}$& $ 5.7 $&   $8^3 \times 48$&          200 \\
B& $ 5.7 $&   $16^3 \times 48$&        100 \\
E${}^{\dagger}$& $ 5.85 $&  $12^3 \times 48$&   100 \\
C${}^{\dagger}$& $ 6.0 $&   $16^3 \times 48$&   100 \\
Q& $ 6.0 $&  $ 12^3 \times 48$& 235  \\
D${}^{\dagger}$& $ 6.3 $&   $24^3 \times 80$&   100 \\
H${}^{\dagger}$& $ 6.52 $&  $ 32^3 \times 100$& 60  \\
\noalign{\smallskip}
\hline
\multicolumn{4}{c}{\mystrut Dynamical, $N_f=2$}\\
\hline
\mystrut  Name& $6/g^2$, $am$& size&\#\\
\hline
F${}^{**}$& $ 5.7$,\ $0.01$&   $16^3 \times 32$&    49  \\
G${}^{\dagger}$${}^{*}$& $ 5.6$,\ $0.01$&   $16^3 \times 32$&    200 \\
L& $ 5.445$,\ $0.025$&  $ 16^3 \times 48$&    100 \\
N& $ 5.5$,\ $0.1$&  $ 24^3 \times 64$&    100 \\
O& $ 5.5$,\ $0.05$&  $ 24^3 \times 64$&    100 \\
M& $ 5.5$,\ $0.025$&  $ 20^3 \times 64$&    100 \\
P& $ 5.5$,\ $0.0125$&  $ 20^3 \times 64$&    100 \\
\hline
\noalign{\medskip}
\noalign{${}^\dagger$  Approximately same physical volume\shortstrut}
\noalign{${}^{**}$  from Columbia group\shortstrut}
\noalign{${}^{*}$ from HEMCGC\shortstrut}
\end{tabular}
\end{table}

There have been five significant changes between the results we 
presented at Lattice 96 \cite{ClaudeMILC96,CraigMILC96}
and those presented here.  We now extrapolate
to a physical $\kappa$ that gives the correct pion mass rather than
to $\kappa_c$.  We now evaluate the perturbative renormalization factors in a 
more consistent way.
We have explored more
fit ranges and this has helped us to refine our estimate of the 
statistical/systematic errors of the fits.  
We have new results for the decay constants in the static
approximation for several additional configuration sets.  
We are now able therefore to determine
physical quantities at the $b$-quark mass, on
essentially all data sets,  
by interpolation between the static limit and propagating quarks
that are lighter than $b$.
(On two data sets, M and P, we still are forced to extrapolate
at the present time; new static results on these sets 
will be completed shortly.) We have also refined our analysis of the
$\pi$ mass and $f_\pi$ so that our chiral fits are better understood
and controlled.
Each of these issues is discussed in a subsequent
subsection.  We then present a summary of our results and, finally, our
conclusions.

\section{CHANGES TO THE ANALYSIS}

\subsection{Chiral extrapolations}
For the results presented at Lattice 96, all our chiral extrapolations
were done to $\kappa_c$.  We now determine the light quark mass
from 
$$ m_\pi(\kappa_l) = m_\pi{}_{\rm phys}\,\,,$$
where $m_\pi{}_{\rm phys}$ is the physical pion mass, and $f_\pi$ is
used to set the scale.  This has resulted in a small reduction of
$f_B$ of 2--6 MeV, for various cases.  Physically, this change comes about
because $a f_\pi^{\rm latt}$ is now bigger, which results in a larger
value for $a$, and hence a smaller value for $f_B$ since $a f_B$ changes
less than $a f_\pi$.

\subsection{Perturbative corrections}
There are several improvements to the way we calculate the perturbative 
corrections to the renormalization constants of the axial-vector current.
Tadpole improvement is applied in a consistent manner\cite{TADPOLE}, 
and the remaining
perturbative corrections are calculated using a boosted coupling constant
$g_V^2(q^*)$.  The scale $q^*$ has been calculated for $Z_A^{\rm stat}$ by
Hernandez and Hill to be $2.18/a$ \cite{HERNANDEZ}.  
For propagating quarks, the value
$1/a$ has been suggested by Lepage and Mackenzie \cite{TADPOLE}
and this value is used in the analysis presented here.  

\begin{table}[tbh]
\caption{Differences (new $-$ old) in values of $f_B$ from current tadpole improvement}
\label{tab:tadpolediffs}
\begin{tabular}{ll}
\hline
\multicolumn{2}{c}{\mystrut Quenched}\\
\hline
\mystrut  $6/g^2$& difference (MeV)\\
\hline
5.7&+5--8\\
5.85&+3\\
6.0&+2\\
6.3&$\pm$1\\
6.52&-0.2\\
\hline 
\multicolumn{2}{c}{\mystrut Dynamical, $N_f=2$}\\ 
\hline
\mystrut $6/g^2$, mass& difference (MeV)\\
\hline
5.445, 0.025&+3\\
5.5, various&4--7\\
5.6, 0.01&+2\\
5.7, 0.01&-0.1\\
\hline
\end{tabular}
\end{table}

Results presented at Lattice 96 followed an approximate method
of Bernard, Labrenz and Soni \cite{BERLABSONI}
that was designed to reproduce the results
of tadpole improvement at $6/g^2=6.3$.  However, that method differs
from the more consistent current method, and the differences increase
with the lattice spacing.  The differences are shown in
Table \ref{tab:tadpolediffs}.  We note however that 
a very recent lattice perturbation theory calculation by Bernard, Golterman 
and McNeile gives $q^*\sim2.3/a$ \cite{BGM} for the $Z_A$ for
Wilson quarks.  When this value
is put into the analysis, the differences listed in Table \ref{tab:tadpolediffs}
will decrease, making the analysis closer to that of Lattice 96.

\begin{figure}[thb]
\epsfxsize=0.99 \hsize
\epsffile{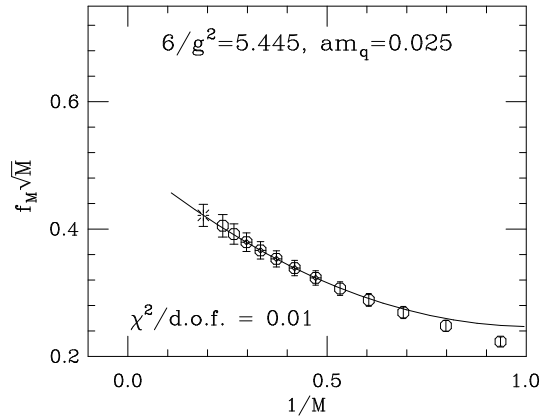}
\caption{$f_M \times M^{1/2}$ {\it vs}. $1/M$ for $6/g^2=5.445$ and $am_q=0.025$ (data set L)
with no static result.}
\label{fig:nostatic5445}
\end{figure}
\begin{figure}[thb]
\vspace{9pt} 
\epsfxsize=0.99 \hsize 
\epsffile{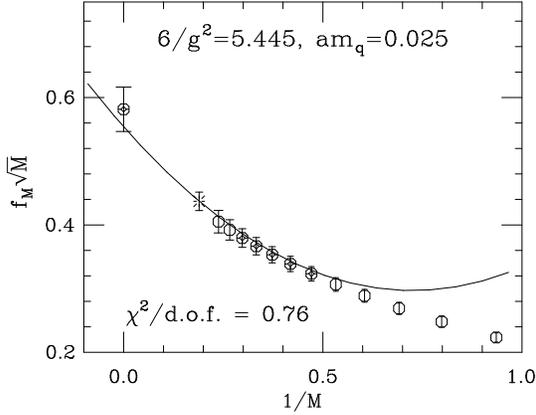} 
\caption{$f_M \times M^{1/2}$ {\it vs}. $1/M$ for $6/g^2=5.445$ and $am_q=0.025$ (data set L)
including the static result.}
\vspace{-18pt}  
\label{fig:static5445}
\end{figure}

\subsection{Alternative fit ranges}
To determine the pseudoscalar mass and decay constant, it is necessary to
fit hadron propagators.  Of course, it is not known {\it a priori}\/ what
range of distances to use in the fit.  It is appropriate to consider the
variation resulting from picking different ranges as part of the
statistical error.  At Lattice 96 we had a limited choice of alternative
ranges used to estimate this error.  We have had the
opportunity to explore additional ranges, and this has changed some of our
statistical errors.  For example, for $6/g^2=6.0$ we had thought that the
variation from changing the range was $\pm15$ MeV.  We currently estimate
it to be $\pm8$ MeV.

\subsection{New static $f_B$ calculations}
All the improvements described above can be implemented after the bulk of
the computation is completed.  However, the calculation of improved values for
$f_B$ in the static approximation involved a reanalysis of many
of our lattices.  In our original calculations, the $\fbstat$ 
results were a
byproduct of the propagating quark calculation.  On large volumes, however,
they were subject to contamination from states with non-zero momentum,
and therefore were left out of the analysis.  On
small volumes the energy gap between the ground state and states with
momentum $2 \pi/L$ was sufficient to suppress the non-zero momentum
states.

Our recent calculations with dynamical quarks
have been on larger volumes than the early ones, and we have done new
calculations to determine the static decay constants.  In some cases, we
also improved the accuracy of the quenched results.  At Lattice 96, Craig
McNeile \cite{CraigMILC96} 
described our method and some of the initial results.  The key
point is that the static quark is smeared relative to the propagating quark
and the basis set includes about 10 different smearing functions.
The analysis of the static $f_B$ data is still being refined.

\begin{figure}[thb]
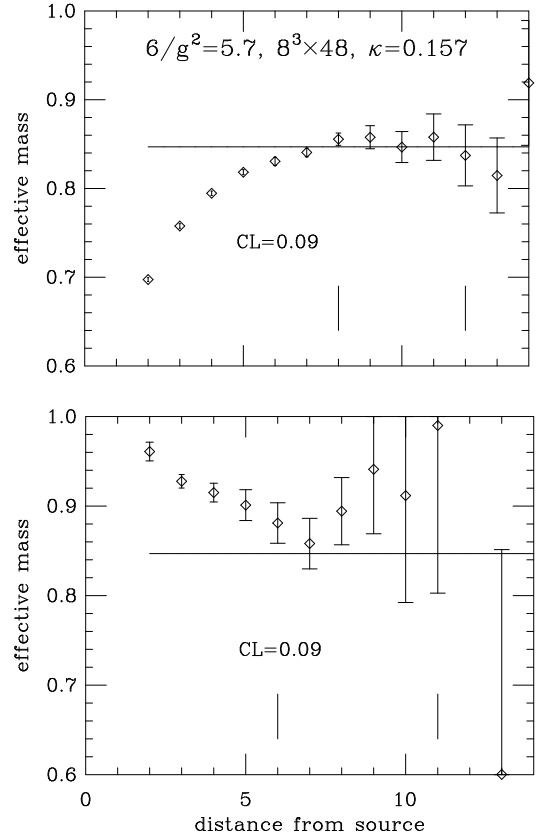

\epsfxsize=0.99 \hsize
\epsffile{oldstatSL157b57.ps}
\vspace{-18pt}  
\epsfxsize=0.99 \hsize 
\epsffile{oldstatSS157b57.ps} 
\vspace{-18pt}  
\caption{Old static results on data set A for effective masses and fitted masses.
(a) smeared-local propagator, (b) smeared-smeared propagator.}
\label{fig:oldstatic57}
\end{figure}

\begin{figure}[thb]
\epsfxsize=0.99 \hsize
\epsffile{newstatSL157b57.ps}
\vspace{-18pt}  
\epsfxsize=0.99 \hsize 
\epsffile{newstatSS157b57.ps} 
\vspace{-18pt}  
\caption{New static results on data set A for effective masses and fitted masses.
(a) smeared-local propagator, (b) smeared-smeared propagator.}
\label{fig:newstatic57}
\end{figure}

\begin{figure}[thb]
\epsfxsize=0.99 \hsize
\epsffile{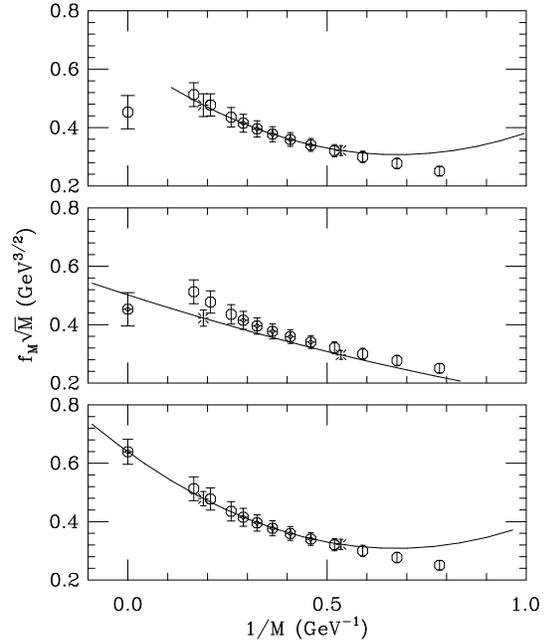}
\vspace{-28pt}
\caption{Decay constant {\it vs}.\ the inverse meson mass on data set A.  
In the top graph (from data analysis presented at Lattice 96) 
the static result is ignored.  
In the middle graph
the old static result is included in the fit.  
In the bottom graph our new static result is included.}
\vspace{-18pt}  
\label{fig:combined57}
\end{figure}

With more accurate results in the static case, we are able to improve some of 
our fits for the decay constant {\it vs}.\ $1/M$, where $M$ is the heavy meson
mass.  Let's consider some examples.

Data set $L$ with $6/g^2=5.445$ and $am_q=0.025$ is a case where we had
no previous static result.  In Fig.~\ref{fig:nostatic5445}, 
we see $f\sqrt M$ plotted {\it vs}.\ $1/M$.  
The curve is fit to the five decorated octagons.  The burst
is the {\em extrapolated} value of the decay constant and corresponds to 203(8)
MeV.  An alternative fit to the points with $1/M>0.5$ gives 200(7)
MeV.  In Fig.~\ref{fig:static5445}, we see that including the 
new static value results in an
{\em interpolated} value of 210(7) for $f_B$.  Using the points to the
right, we find $f_B=209(7)$ MeV.  The results for the two choices of
fitting range are more consistent and the error is slightly reduced.

As another example, for case G with
$6/g^2=5.6$ and $am_q=0.01$ our old estimate of
$f_B^{static}$ was 265(12)(10), where the first error is statistical for a
fixed fit range and the second is the variation from the fit range.  
Our new result is 285(4)(1).  The errors are considerably
reduced and the result is somewhat higher.

Finally, we consider a quenched example, case $A$.  In this particular
case, although
the volume is small enough to avoid contributions from higher-momentum
states, we had poor plateaus in our effective mass plots.  Figures
\ref{fig:oldstatic57} and \ref{fig:newstatic57} 
show effective mass plots for the old calculation and the new
static calculations.  In each case we show two effective mass plots, one
for the smeared source and local sink and one for smeared source and sink.
The vertical bars near the bottom of the graph show the fitting range.
The horizontal lines give the fitted mass.  The confidence level of a fit
including the old static result was just 0.09.  With the new results it is 0.7.
The new results approach the asymptotic value more quickly and the errors 
in the effective masses are smaller.  The old value for $\fbstat$ was 216(27)
MeV; the new value, 283(18).
Using the old static value  to determine $\fb$, we would
have gotten 201(13) MeV.  Without the
static value included in the fit, we obtained 227(18) MeV. (This
was the fit chosen at the time of Lattice 96.)  Including our
new static value gives 229(12) MeV, in agreement with the 
Lattice 96 result, but with a smaller error.
In Fig.~\ref{fig:combined57}, we see all three fits.  The points determining
the curve are the decorated octagons.  The bursts show both $\fb$ and
$f_D$.  [Note: we quote statistical errors only for $\fb$ here.
Choices of different fitting ranges result in additional errors of
order 15 MeV.  However, the relations among the various treatments of the
static point remain essentially unchanged.]

\begin{figure}[thb]
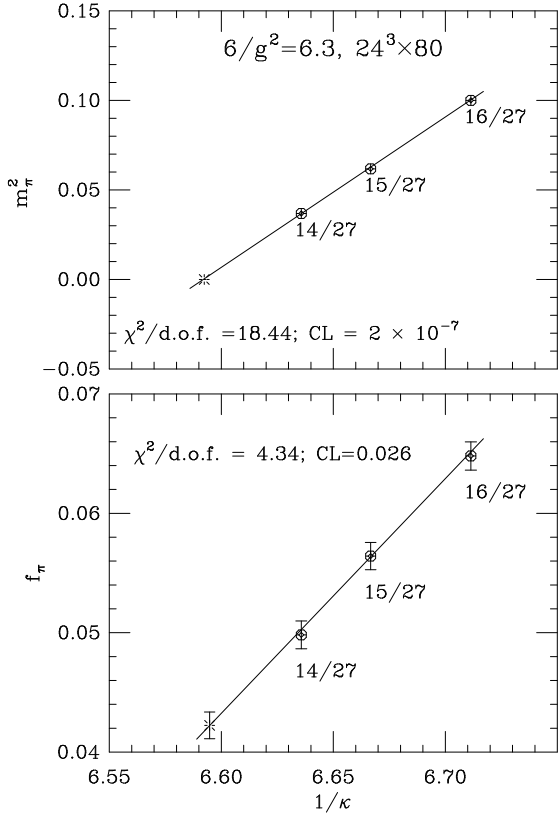

\epsfxsize=0.99 \hsize
\epsffile{mpisqrfixed.ps}
\vspace{-28pt}
\epsfxsize=0.99 \hsize
\epsffile{fpifixed.ps}
\vspace{-18pt}  
\caption{Chiral extrapolation for $m_\pi^2$ and $f_\pi$ on data set D
with a fixed fitting range}
\label{fig:fixed}
\end{figure}

\subsection{Scale determination}
\begin{figure}[thb]
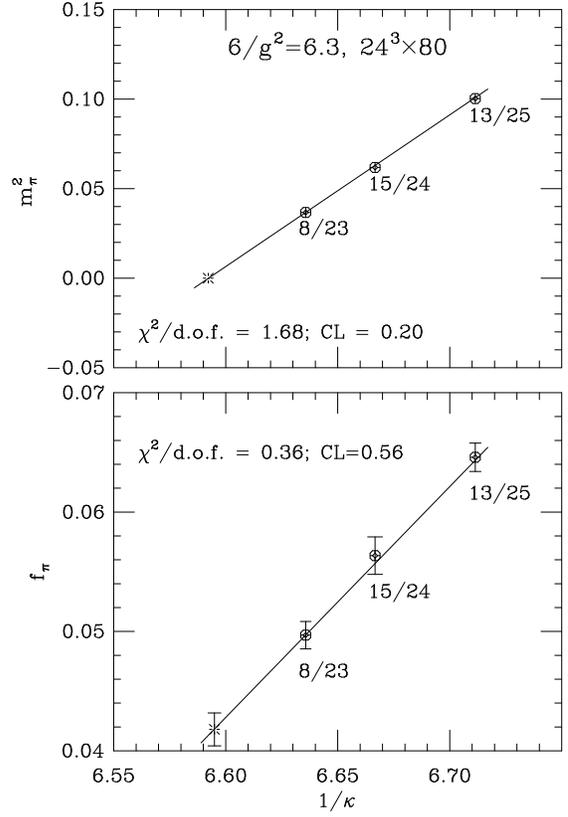

\epsfxsize=0.99 \hsize
\epsffile{mpisqrsliding.ps}
\vspace{-28pt}
\epsfxsize=0.99 \hsize
\epsffile{fpisliding.ps}
\vspace{-18pt}  
\caption{Chiral extrapolation for $m_\pi^2$ and $f_\pi$ on data set D
with different fitting ranges}
\label{fig:sliding}
\end{figure}

The final improvement we have made to our analysis involves the
determination of the lattice spacing from $f_\pi$.  As mentioned at Lattice
96, we often had a difficult time getting good linear fits for $m_\pi^2$ or
$f_\pi$ {\it vs}.\  $1/\kappa$.  
We used quadratic fits to estimate the systematic error in the
scale, resulting in variations from 8 to 25\%.  However, our earlier fits for
$m_\pi^2$ and $f_\pi$ were based upon a fixed fitting range for all three
light $\kappa$ values.  We have relaxed this constraint and let the fitting
range vary.  This has certainly helped to improve the confidence level 
of the chiral extrapolation.  As an example, for case D, we show
both $m_\pi^2$ and $f_\pi$ in Figs.~\ref{fig:fixed} and
\ref{fig:sliding}.  The individual points in the plot
are labeled with $\chi^2$/(degrees of freedom) for the fits to the
pion propagators as a function of distance from source.  We see, 
particularly at the lightest $\kappa$, that we are able to improve the fit by
adjusting the range.  The confidence level of the linear chiral
fits is determined by the $\chi^2$ of the full covariance matrix.
For the fixed fit ranges, we have confidence levels of $2\times10^{-7}$ and
0.026 for linear fits to $m_\pi^2$ and $f_\pi$, respectively.
When we allow the fitting range to vary, the
confidence levels improve to 0.20 and 0.56, for the corresponding cases.
We have applied the same procedure to case O, and the linear fits in
$1/\kappa$ are improved, but not yet good.  We have not yet implemented
this for every case.

We have both case A and B in order to study finite volume effects on
$\fb$; however, $f_\pi$ also varies.  We have recently completed production
running on a large scale finite volume study of the light quark spectrum 
at the same gauge coupling \cite{SGSPEC}.  
We studied six quark masses on four lattice sizes
from $N_s=12$ to 24.  We should be able to study both the linearity of the
chiral extrapolation and finite volume effects to higher accuracy.

\begin{figure}[thb]
\epsfxsize=0.99 \hsize
\epsffile{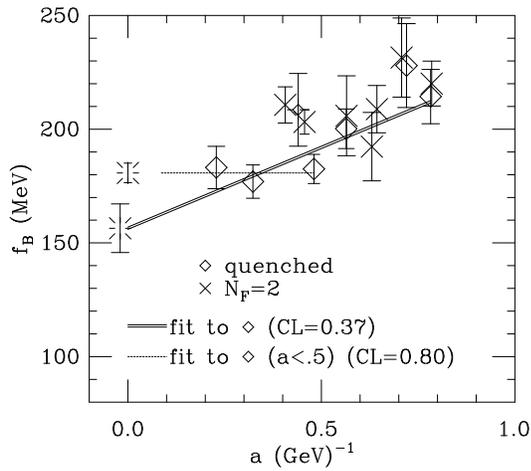}
\vspace{-28pt}
\caption{$\fb$ {\it vs}.\ lattice spacing.}
\label{fig:fb}
\end{figure}
\begin{figure}[thb]
\epsfxsize=0.99 \hsize
\epsffile{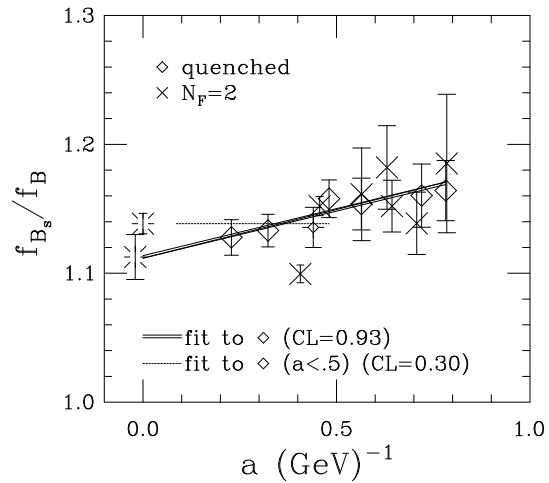}
\vspace{-28pt}
\caption{$f_{B_s}/f_B$ {\it vs}.\ lattice spacing.}
\label{fig:fbsoverfb}
\end{figure}

\section{RESULTS}
Having carefully fit the hadron propagators and done the mass
extrapolations, we are now ready to address the lattice spacing dependence
of our results.  We have results for $f_B$ ($f_D$) and 
$f_{B_s}$ ($f_{D_s}$) and their ratios.  Figure~\ref{fig:fb}
shows our latest
results for $f_B$.  The diamonds are the quenched results and two fits
are shown for their $a$ dependence.  The horizontal line is a constant fit to
the three smallest lattice spacings.  There is also a linear fit to all of
the diamonds.  Both fits have quite reasonable confidence levels.
The bursts plotted near $a=0$ are the extrapolated values.  
In Table~\ref{tab:results}, we
list central values coming from the linear fit.  The errors in the graphs
are based upon statistical and fitting errors, and are the first errors
listed in the table.  The second error is our estimate of the 
systematic errors within the quenched approximation.  The chief error here
is that from the ambiguity in how to extrapolate in $a$. 
Other sources of error include the
chiral extrapolation, finite volume effects, weak coupling perturbation
theory, the extrapolation in $1/M$ and large $ma$ effects.  How we deal
with each of these effects has been discussed in Ref.~\cite{US}.  
We must admit
that we are wavering between taking our central value from the
constant and the linear fit to the lattice spacing dependence.
Since we are currently using the value from the linear fit, the second
error is more likely to be positive than negative.
The error from using the quenched approximation we consider to be rough at
present.  We now have results for seven sets of dynamical quark
parameters.  They are denoted by crosses in the figures.
When we have finished our calculations with coupling 5.6, we will attempt an
extrapolation in the dynamical quark mass.

In Fig.~\ref{fig:fbsoverfb}, we show the ratio $f_{B_s}/\fb$ as a function
of lattice spacing.  One expects that in a ratio many of the
systematic errors will be common to the two quantities
and that the ratio can be calculated more accurately than either quantity.  
Here there is better agreement between the constant fit to the three weakest
couplings and the linear fit to all the quenched couplings.  There is one
point that seems to stand out from the others.  This is case G, an analysis
of $16^3\times32$, $6/g^2=5.7$ dynamical quark lattices that were provided to
us by the Columbia group \cite{COLUMBIA}.  
These lattices are physically the smallest
lattices we have studied.  We tried to test whether we might be seeing a
finite size effect by doing a new quenched calculation on a small volume.
Our most comparable quenched coupling is 6.0 where we studied $16^3\times48$
lattices, so we reduced the spatial size to 12 for this calculation
(lattice Q).
This smaller volume calculation is plotted in Figs.~\ref{fig:fb} and
\ref{fig:fbsoverfb} with the fancy
diamond.  Although it appears a little lower than the larger quenched
volume, the effect is not as pronounced as for the Columbia lattices.
Thus, we do not have a credible explanation of this as a systematic
effect.

Of particular note among our results is the value of $f_{D_s}$.  The
world average
experimental result has recently been reported in a review
talk by J.~Richman \cite{RICHMAN}.  The result, $241\pm21\pm30$ MeV compares 
quite reasonably with ours.

\begin{table}
\caption{Preliminary results for meson decay constants}
\label{tab:results}
\begin{eqnarray*}
\hline
\mystrut f_B&= 156\ \pm 11\ \pm 30\ \pm 14\ \ \ {\rm MeV}\\
f_{B_s}&= 173\ \pm 9\ \pm 40\ \pm 18\ \ \ {\rm MeV}\\
f_{B_s}/f_B&= 1.11\ \pm 0.02\ \pm 0.03\ \pm 0.07  \\
f_D&= 192\ \pm 9\ \pm 18\ \pm 10\ \ \ {\rm MeV}\\
f_{D_s}&= 209 \pm 7\ \pm 27\ \pm 12\ \ \  {\rm MeV}\\
f_{D_s}/f_D&= 1.09\ \pm 0.02\ \pm 0.05\ \pm 0.05\ \\
\hline
\end{eqnarray*}
\end{table}

\section{CONCLUSIONS }

We feel that our calculations of the meson decay constants have progressed 
to the stage that we can control the sources of systematic error 
within the quenched approximation.  At this point, the major source of
uncertainty within that approximation is whether to use a linear
extrapolation for all of our couplings or a constant fit to the weakest
ones, to extract the continuum results.

Our calculations with dynamical quarks are now giving us
a rough bound on the error from quenching, but we are not yet in a position to
extrapolate in dynamical quark mass or lattice spacing.  We are still
running on lattices with $6/g^2=5.6$.
Beyond these calculations, one can imagine putting in a dynamical
strange quark, or repeating this calculation with dynamical Wilson quarks.
Such ambitious plans are not on our agenda at the present time.

The apparent agreement between lattice calculations of
\fds \ and the recent experimental results is encouraging,
but the errors on both are still rather large.   With 
new experimental facilities
in Japan and the United States coming on-line in the near future, 
and with continuing improvement in the lattice results, we expect
increasing interplay between experimentalists and the lattice community.

\section*{ACKNOWLEDGEMENTS}
This work was supported by the U.S. Department of Energy under contracts
DE-AC02-76CH-0016,
DE-AC02-86ER-40253,
DE-FG03-95ER-40906,
DE-FG05-85ER250000,
DE-FG05-96ER40979,
DE-2FG02-91ER-40628,
DE-FG02-91ER-40661,
and National Science Foundation grants
NSF-PHY93-09458,
NSF-PHY96-01227,
NSF-PHY91-16964.
Computations were performed at the Oak Ridge National Laboratory
Center for Computational Sciences, the Pittsburgh
Supercomputing Center,  the National Center for Supercomputing
Applications, the Cornell Theory Center, the San Diego Supercomputer
Center, and Indiana University.
Two of us, (S.G. and D.T.), are very grateful to the Center for
Computational Physics at the University of Tsukuba for its generous
support and warm hospitality.  In addition, we thank Profs.~Y.~Iwasaki and
A.~Ukawa for all their efforts in organizing a very stimulating workshop.

\end{document}